
%
%
\magnification=\magstep1
\def\ssp{\baselineskip=11pt plus 1pt minus 1pt}

\spaceskip=0.4em plus 0.15em minus 0.15em
\xspaceskip=0.5em
\hsize=15 true cm
\hoffset=0 true cm
\vsize=22 true cm

\def\ref{\par\noindent\hangindent 20pt}
\def\sles{\lower2pt\hbox{$\buildrel {\scriptstyle <}
   \over {\scriptstyle\sim}$}}

\def\sgreat{\lower2pt\hbox{$\buildrel {\scriptstyle >}
   \over {\scriptstyle\sim}$}}

\def\lapprox{\lower2pt\hbox{$\buildrel \lower2pt\hbox{${\scriptstyle<}$}
   \over {\scriptstyle\approx}$}}

\def\gapprox{\lower2pt\hbox{$\buildrel \lower2pt\hbox{${\scriptstyle>}$}
   \over {\scriptstyle\approx}$}}

\def\both{\lower2pt \hbox{$\buildrel {\leftarrow} \over {\rightarrow}$}}


\ssp
\centerline{\bf Advection-Dominated Accretion: Self-Similarity and
Bipolar Outflows}
\bigskip
\centerline{Ramesh Narayan and Insu Yi}
\bigskip
\centerline{Harvard-Smithsonian Center for Astrophysics}
\centerline{60 Garden Street, Cambridge, MA 02138}
\vskip .4in

\noindent{\bf Abstract}
\bigskip
We consider axisymmetric viscous accretion flows where a fraction $f$
of the viscously dissipated energy is stored in the accreting gas as
entropy and a fraction $1-f$ is radiated.  Assuming $\alpha$-viscosity
we obtain a two-parameter family of self-similar solutions.  Very few
such exact self-consistent solutions are known for viscous
differentially rotating flows.  When the parameter $f$ is small, that
is when there is very little advection, our solutions resemble
standard thin accretion disks in many respects except that they have a
hot tenuous corona above the disk.  In the opposite {\it
advection-dominated} limit, when $f\rightarrow1$, the solutions
approach nearly spherical accretion.  The gas is almost at virial
temperature, rotates at much below the Keplerian rate, and the flow is
much more akin to Bondi accretion than to disk accretion.  None of the
solutions have funnels.

We compare our exact self-similar solutions with approximate solutions
which had been previously obtained using a height-integrated system of
equations.  We find that various dynamical variables such as the
radial velocity, angular velocity and sound speed estimated from the
approximate solutions agree very well with the corresponding
spherically averaged quantities in the exact solutions.  We conclude
that the height-integration approximation is an excellent one for a
wide range of accretion conditions, including nearly spherical flows,
provided the equations are interpreted as spherical averages.

We find that the Bernoulli parameter is positive in all our solutions,
especially close to the rotation axis.  This effect is produced by
viscous transport of energy from small to large radii and from the
equator to the poles.  In addition, all the solutions are convectively
unstable and the convection is especially important near the rotation
axis.  For both reasons, we suggest that a bipolar outflow will
develop along the axis of these flows, fed by material from the
surface layers of the equatorial inflow.

\bigskip\bigskip\noindent
Subject headings: accretion, accretion disks: black hole physics: hydrodynamics

\vfill\eject
\noindent
{\bf 1.  Introduction}
\bigskip
In a previous paper (Narayan \& Yi 1994, hereafter NY) we discussed
the potential importance of advection effects in accretion flows.  We
showed that, both at very low and very high optical depths, the energy
released through viscous stresses in an accretion disk may be trapped
within the accreting gas.  Most of the energy then is advected with
the flow as stored entropy.  Such {\it advection-dominated} flows have
been found in models of boundary layers in cataclysmic variables at
low accretion rates (Narayan \& Popham 1993) and in models of pre-main
sequence stars such as the FU Orionis systems at very high accretion
rates (Popham et al 1993).  Advection-dominated conditions may also
occur in the inner parts of disks around neutron stars and black
holes.

In the analysis presented in NY, we integrated the flow equations in
the ``vertical'' direction, i.e. parallel to the rotation axis.
Making the usual assumptions of steady state, axisymmetry, and
$\alpha$-viscosity, we obtained a set of ordinary differential
equations for the gas variables as a function of the cylindrical
radius $R$.  We showed that these equations have an exact self-similar
solution where all variables have power-law dependences on $R$ and
where the Mach number is independent of $R$.  This solution, which had
previously been discovered by Spruit et al.  (1987), has several
interesting and unexpected properties.  However, before we can explore
the consequences of these properties we need first to confirm that the
self-similar solution itself is real and not just an artifact of the
vertical integration of the equations.

Vertical integration is a standard approximation which has been used
in accretion disk studies from the earliest days.  The physical
motivation behind this approximation is that the vertical thickness of
an accretion disk is usually much smaller than the local radius, so
that the flow velocities are likely to be more or less independent of
height.  It is then reasonable to expect that very little is lost by
integrating out the vertical coordinate.  Unfortunately, the
self-similar solutions discovered by Spruit et al. (1987) and NY are
not thin.  The temperature of the accreting gas is nearly always close
to virial, and the formal vertical thickness is comparable to the
radius.  This inconsistency raises serious questions concerning the
validity of the solutions and the reliability of the conclusions.

In this paper, we avoid the height-integration approximation and
instead set up exact flow equations for steady axisymmetric flow
in the $r\theta$ plane.  The equations we obtain are similar
to those written down by Begelman \& Meier (1982).  We present here
numerical self-similar solutions of the equations.  These solutions
are direct generalizations of the height-integrated solutions in NY.
Rather gratifyingly, we find that all the features of the
height-integrated solution are reproduced well in the present
solutions.  In fact, we even find excellent {\it quantitative}
agreement between the two approaches, suggesting that the
height-integration approximation may be much better than previously
thought.

In \S 2 we write down the equations that we solve along with the
boundary conditions.  Then in \S 3 we describe numerical results of
the equations, discuss some of the more interesting properties of the
solutions, and make various comparisons with the previously-obtained
height-integrated solutions.  Finally, in \S 4 we discuss the
implications of the results, especially for the formation of outflows.
Appendix A compares our solutions with those obtained by Begelman \&
Meier (1982), and Appendix B discusses the relationship to Bondi
(1952) accretion.

\bigskip\bigskip\noindent
{\bf 2.  Equations of a Steady Axisymetric Advection-Dominated Flow}
\bigskip
We define the isothermal sound speed $c_s$ by
$$
p = \rho c_s^2, \eqno(2.1)
$$
where $p$ is the pressure and $\rho$ is the density, and we assume
that the gas has a fixed ratio of specific heats, $\gamma \equiv c_p /
c_v$.  For convenience we define
$$
\epsilon = {5/3 - \gamma \over \gamma -1}.\eqno(2.2)
$$
We work in spherical polar coordinates $r\theta \phi$, and write the
three components of velocity as $v_r,~ v_{\theta}$ and $v_{\phi} =
\Omega r \sin \theta$, where $\Omega$ is the angular velocity.  The
gas accretes onto a central mass $M$ and we define the Keplerian
angular velocity $\Omega _K(r)$ by
$$
\Omega_K(r) = \left(GM \over r^3 \right)^{1/2}. \eqno(2.3)
$$
Assuming steady state $(\partial/\partial t = 0)$ and axisymmetry
$(\partial/\partial \phi = 0)$, the continuity equation gives
$$
{1\over r^2}{\partial \over \partial r} (r^2\rho v_r) + {1\over r}
{\partial \over \partial \theta} (\rho v_{\theta}) = 0.\eqno(2.4)
$$
Integrating this over angle we obtain the net mass accretion rate,
$$
\dot M = - \int 2\pi r^2 \sin\theta \rho v_r d\theta. \eqno(2.5)
$$

We employ the usual $\alpha$-prescription for the viscosity, which we
write in the following form for the kinematic coefficient of
viscosity,
$$
\nu = {\alpha c^2_s \over \Omega_K}, \eqno(2.6)
$$
where $\alpha$ is a constant.  This form is equivalent to the
assumption that $\nu\sim\alpha c_sH$ where $H\sim c_s/\Omega_K$ is the
``vertical'' scale height.  Note that $\nu$ is a function of position,
both because $\Omega_K$ depends on $r$ and because $c_s$ varies from
point to point.  The three components of the momentum equation give
(e.g., Mihalas \& Mihalas 1984)
$$
\rho\left(v_r{\partial v_r\over \partial r}-{v_{\phi}^2\over r}\right)
=-{GM\rho\over r^2}-{\partial p\over \partial r}
+{\partial\over\partial r}\left[2\nu\rho{\partial v_r\over \partial r}
-{2\over 3}\nu\rho\left({2v_r\over r}+{\partial v_r\over \partial r}\right)
\right]
$$
$$
\qquad\qquad\qquad\qquad\qquad\qquad
+{1\over r}{\partial\over \partial\theta}\left({\nu\rho\over r}{\partial v_r
\over\partial\theta}\right)
+{\nu\rho\over r}\left[4r{\partial\over \partial r}\left(v_r\over r\right)+
{\cot\theta\over r}{\partial v_r\over \partial \theta}\right],\eqno(2.7)
$$

$$
\rho\left(-{\cot\theta\over r} v_{\phi}^2 \right)
=-{1\over r}{\partial p\over \partial\theta}
+{\partial\over \partial r}\left({\nu\rho\over r}{\partial v_r\over\partial
\theta}\right)\qquad\qquad\qquad\qquad\qquad
$$

$$
\qquad\qquad\qquad
+{1\over r}{\partial\over\partial\theta}\left[{2\nu\rho v_r\over r}
-{2\nu\rho\over 3r}
\left({2v_r\over r}+{\partial v_r\over \partial r}\right)\right]
+{3\nu\rho\over r^2}{\partial v_r\over \partial\theta},\eqno(2.8)
$$

$$
\rho\left(v_r{\partial v_{\phi}\over \partial r}+{v_{\phi} v_r\over r}\right)=
{\partial\over \partial r}\left[\nu\rho r{\partial\over \partial r}
\left(v_{\phi}\over r\right)\right]+{1\over r}{\partial\over \partial\theta}
\left[{\nu\rho\sin\theta\over
r}{\partial\over\partial\theta}\left(v_{\phi}\over
\sin\theta\right)\right]
$$
$$
\qquad\qquad\qquad\qquad\qquad\qquad
+{\nu\rho\over r}\left[3r{\partial\over\partial r}\left(v_{\phi}\over r\right)
+{2\cot\theta\sin\theta\over r}{\partial\over\partial\theta}\left(v_{\phi}\over
\sin\theta\right)\right],\eqno (2.9)
$$
while the energy equation gives
$$\rho\left(v_r{\partial e\over\partial r}-{p\over \rho^2}
v_r{\partial\rho\over
\partial r}\right)
=
-{2f\nu\rho\over 3}\left[{1\over r^2}{\partial\over\partial r}
\left(r^2 v_r\right)\right]^2 + 2f\nu\rho\times\qquad\qquad\qquad\qquad\qquad
$$
$$
\left[\left(\partial v_r\over\partial r\right)^2
+2\left(v_r\over r\right)^2+{1\over 2}\left({1\over r}{\partial v_r\over
\partial \theta}\right)^2
+{1\over 2}\left(r{\partial\over \partial r}
\left(v_r\over r\right)\right)^2
+{1\over 2}\left({\sin\theta\over r}{\partial\over\partial\theta}\left(v_{\phi}
\over\sin\theta\right)\right)^2
\right].\eqno(2.10)
$$
Anticipating the self-similarity form assumed below (eqs 2.11--2.15),
we have set $v_\theta=0$ in the above equations.  The left-hand side
of equation (2.10) is the gradient of the entropy.  The right-hand
side is the rate of generation of energy through viscous dissipation,
except that it is multiplied by a parameter $f$.  This parameter
describes the fraction of the disipated energy which is advected as
stored entropy, and is therefore a measure of the degree to which the
flow is advection-dominated.  A fraction $(1 - f)$ of the energy is
removed through radiative losses.  In principle, $f$ could be a
function of $\theta$, but all the results we present here correspond
to the simplest assumption, viz. $f =$ constant.

\medskip
We restrict ourselves in this paper to self-similar flows.  We
therefore seek a solution of the form
$$
\eqalignno{
\rho & = r^{-3/2} \rho(\theta), & (2.11) \cr
v_r & = \sqrt{GM\over r} v(\theta) = r\Omega_K(r)v(\theta), & (2.12) \cr
v_{\theta} & = 0, & (2.13) \cr
v_{\phi} & = r\Omega_K (r) \Omega (\theta), & (2.14) \cr
c_s & = r\Omega_K (r) c_s(\theta). & (2.15)
}
$$
The form of the solution is obvious.  The only lengthscale in the
problem is $r$ and the only frequency is $\Omega_K$.  Therefore, all
velocities must scale with radius as $r\Omega_K$.  Angular variations
at a given radius are modeled through the dimensionless functions
$v(\theta), ~\Omega(\theta)$ and $c_s(\theta)$.  Given the radial
scaling of $v_r$, the scaling of $\rho$ is uniquely determined by the
constancy of $\dot M$ in (2.5), and since $r^2\rho v_r$ is independent
of $r$, the continuity equation (2.4) shows that $v_\theta=0$.

By construction, the solution (2.11) -- (2.15) automatically satisfies
the continuity eq. (2.4). Substituting the solution in the momentum
and energy equations we obtain the following four coupled differential
equations in $\theta$:
$$
\eqalignno{
- {1\over 2}v^2 - \sin^2\theta \Omega^2 & = -1+c^2_s\left({5\over 2} -
\alpha v + \alpha \cot \theta {dv\over d \theta} \right) + {1\over \rho}
{d\over d\theta}\left( \alpha \rho c^2_s {dv\over d\theta}\right), & (2.16) \cr
-\cos \theta \sin \theta \Omega^2 & = - {1\over \rho} {d\over d\theta}
(\rho c^2_s) + {\alpha c_s^2\over 2}{dv\over d\theta} + {1\over \rho}
{d\over d\theta} (\alpha c^2_s \rho v), & (2.17) \cr
{1\over 2} \sin \theta v \Omega & = - {3 \alpha \sin \theta c^2_s \Omega
\over 4} + {1 \over \rho} {d \over d\theta} \left( \alpha \sin \theta
\rho c^2_s {d\Omega \over d\theta}\right) + 2 \alpha \cos \theta
c^2_s {d\Omega \over d\theta}, & (2.18) \cr
-{3 \epsilon^{\prime} v \over 2 \alpha} & = 3 v^2 + {9 \over 4} \sin^2
\theta \Omega^2 + \sin^2\theta \left( {d\Omega \over d\theta} \right)^2
+ \left( {dv\over d\theta}\right)^2. & (2.19)
}
$$
Following NY, we have introduced in eq. (2.19) a quantity
$\epsilon^{\prime}$ which we define by
$$
\epsilon^{\prime} = {\epsilon \over f} = {1\over f} \left({5/3 - \gamma \over
\gamma - 1}\right), \eqno(2.20)
$$
where $f$ is the parameter we have already introduced in eq. (2.10).

Equations (2.16) -- (2.20) constitute a sixth-order system of ordinary
differential equations for the four functions $v(\theta),~
\Omega(\theta),~ c_s(\theta)$, and $\rho(\theta)$.
The integral (2.5) sets the normalization of $\rho(\theta)$ and
provides one boundary condition.  The remaining boundary conditions
are distributed between the equatorial plane, $\theta = \pi /2$, and
the rotation axis, $\theta = 0$.  At $\theta = \pi /2$, we have by
symmetry the conditions
$$
\theta = {\pi\over 2}:\qquad {dv\over d\theta} = {d\Omega \over d\theta} =
{dc_s \over d\theta} = {d\rho \over d\theta} = 0. \eqno(2.21)
$$
At $\theta = 0$, we insist that the solutions be well-behaved and
non-singular.  This leads to the conditions
$$
\theta = 0:\qquad {d \Omega \over d \theta} = {d v \over
d \theta} = {d c_s \over d \theta} = {d \rho
\over d \theta} = 0, ~ v = 0. \eqno(2.22)
$$
The last condition follows from eq. (2.19).  Not all of the conditions
(2.21), (2.22) are independent.  We choose a convenient subset of
these conditions and solve the differential equations (2.16)--(2.19)
using a numerical relaxation technique (e.g., Press et al. 1992).

Under certain conditions, the sixth-order set of equations
(2.16)--(2.19) reduces to a second-order system.  We discuss this
simplification in Appendix A and compare our work to a previous
analysis of this problem by Begelman \& Meier (1982)

Technically, equation (2.19) allows two possible boundary conditions
on $v$ at $\theta=0$, viz. $v=0$ which is the one we have given in eq
(2.22) and $v=-\epsilon^{\prime}/2\alpha$.  We have tried to obtain
rotating solutions which satisfy the second condition and have been
unable to find any.  However, the boundary condition
$v=-\epsilon'/2\alpha$ does allow a purely spherical non-rotating
inflow solution which we discuss in Appendix B.  This solution is a
generalization of Bondi flow to the case when there is viscosity.

\bigskip\bigskip
\noindent
{\bf 3.  Results}
\bigskip
\noindent
{\bf 3.1  Typical Solutions}
\bigskip
We have obtained numerical solutions of eqs. (2.16) -- (2.19) for a
variety of values of the viscosity parameter $\alpha$ and the
thermodynamic parameter $\epsilon^{\prime}$ (defined in eqs. (2.6) and
(2.20)).  Figure 1 shows a typical sequence of solutions corresponding
to $\alpha = 0.1$ and $\epsilon^{\prime} = 0.1, ~ 1, ~ 10$.  These
solutions may be considered either as a sequence of fully
advection-dominated flows ($f=1$) with $\gamma = 1.6061, ~ 1.3333, ~
1.0606$, or as flows with a fixed value of $\gamma$ and with a
sequence of decreasing $f$ or increasing cooling.

The four panels in Fig. 1 show the variation with polar angle $\theta$
of various dynamical quantities in the solutions.  The top left panel
displays the dimensionless angular velocity $\Omega(\theta)$.  Rather
surprisingly, we find that $\Omega$ is nearly independent of $\theta$
in each solution, varying by only $\sim10\%$ from $\theta = 0$ to
$\theta =\pi /2$.  Radial shells therefore rotate more-or-less
rigidly, but of course there is differential rotation between
neighboring shells.  The actual value of $\Omega$ varies significantly
from one solution to another, changing from $\Omega \sim 0.2$ at
$\epsilon^ {\prime} = 0.1$ to $\Omega \sim 0.9$ at $\epsilon^{\prime}
= 10$.  The scaling of $\Omega$ with $\epsilon'$ follows eq (3.2).
Note that $\Omega\propto(\epsilon')^{1/2}$ for $\epsilon'\ll1$.  This
implies that $\Omega\rightarrow0$ as $\gamma\rightarrow5/3$.

The top right panel shows the radial velocity profiles $v(\theta)$ of
the solutions.  The velocity is zero at $\theta = 0$ (this is a
boundary condition) and maximum at $\theta = \pi /2$.  We find that
$v$ is essentially independent of $\epsilon^{\prime}$ for
$\epsilon^{\prime}\ll1$ and varies as $v\propto
1/\sqrt{\epsilon^{\prime}}$ for $\epsilon^{\prime}\gg1$ (see eq 3.1).

The bottom left panel shows profiles of the density $\rho(\theta)$.
In the $\epsilon^{\prime} = 0.1$ solution $\rho$ varies by only $\sim
10\%$ from $\theta = 0$ to $\theta = \pi /2$.  The solution therefore
corresponds to a nearly spherical configuration.  This is demonstrated
in Figure 2 (top left) where we display isodensity contours in the
meridional plane.  The resemblance of this solution to a star is
striking, but of course it is not a normal star, since it involves a
steady accretion flow.  In any case, it is quite clear that this
solution is very definitely not a ``disk'' in the usual sense.  The
density contrast between $\rho(0)$ and $\rho(\pi /2)$ increases with
increasing $\epsilon^{\prime}$, becoming a factor $\sim 2$ at
$\epsilon^{\prime} = 1$, and a factor $\sim 50$ at $\epsilon^{\prime}
= 10$.  The isodensity contours of these solutions are shown in the
top right and bottom left panels of Fig. 2.  The $\epsilon^{\prime} =
1$ solution looks like a rotationally flattened star, while the
$\epsilon^{\prime} = 10$ solution is beginning to resemble a standard
thin disk.  A value of $\epsilon^{\prime} = 10$ normally implies a
small value of $f$ (unless $\gamma$ is close to 1), and so the
solution represents a case where there is significant cooling.  This
is precisely the limit where we expect the flow to occur in a thin
disk.

The $v(\theta)$ and $\rho(\theta)$ profiles both peak at $\theta = \pi
/2$.  Therefore, in all our solutions the bulk of the accretion occurs
along the equatorial plane, and the accretion rate goes to zero along
the rotation pole.

The bottom right panel of Fig. 1 shows the variation of $c^2_s$, or
equivalently the gas temperature, with $\theta$.  In the
$\epsilon^{\prime} = 0.1$ solution, $c^2_s$ is almost independent of
$\theta$, and the pressure $p =\rho c^2_s$ too is independent of
$\theta$.  In this solution, the rotation is highly sub-Keplerian and
so hydrostatic equilibrium requires primarily a balance between
gravity and the pressue gradient.  Since gravity acts in the radial
direction, the pressure gradient too is almost radial.  The
$\epsilon^{\prime} = 1$ and especially the $\epsilon^{\prime} = 10$
solutions have larger temperature and pressure variations with
$\theta$ and have non-radial pressure gradients.  This is to be
expected given the more rapid rotation of these solutions and the
increasing importance of centrifugal acceleration.

An interesting feature of the large $\epsilon^{\prime}$ solutions is
worth emphasizing.  As already mentioned, these solutions have
efficient cooling and therefore resemble thin disks.  Nevertheless, in
all cases there is a low density {\it corona} above the disk which is
at nearly virial temperature.  This is illustrated by the
$\epsilon^{\prime}=10$ solution in Fig. 1.  The hottest temperature is
achieved at the rotation poles, $\theta=0,~\pi$.

In addition to the above examples with $\alpha = 0.1$, we have
calculated a number of solutions with other values of $\alpha$.  For
$\alpha\ll 1$, $v$ scales as $\alpha$, but except for this, solutions
with the same $\epsilon^{\prime}$ but with different values of
$\alpha$ are virtually indistinguishable from one another.  There are
more significant variations when $\alpha$ exceeds unity.  However,
such large values of $\alpha$ are probably unlikely (e.g.  Narayan,
Loeb \& Kumar 1994, Hawley, Gammie \& Balbus 1994), and we have not
explored this region of parameter space.

In the advection-dominated limit, where $\epsilon'\ll1$, our solutions
have very simple scalings, viz. $v\sim-\alpha$, $\Omega\sim(\epsilon')
^{1/2}$, $c_s\sim1$.  These scalings arise as follows.  Since the
cooling is inefficient, all the viscous energy is stored in the
accreting gas, and this means that the thermal velocities approach
virial speeds, i.e $c_s\sim1$.  Comparing the various terms in eq
(2.18) we see that $v$ has to scale as $-\alpha c_s^2$.  This gives
$v\sim-\alpha$ for a virial gas.  The physical reason for the scaling
is that the radial velocity in an accretion flow is determined
primarily by the rate at which angular momentum is removed from the
gas and this depends on the viscosity coefficient $\nu$.  In the
$\alpha$ prescription (eq 2.6), we have $\nu\sim\alpha$ in scaled
units and this therefore implies $v\sim-\alpha$.  Finally, the scaling
of $\Omega$ arises through the energy equation (2.19).  The left-hand
side of this equation is the product of the radial gradient of the
entropy and the radial velocity, and therefore represents the steady
state rate of change of entropy of a parcel of accreting gas.  Recall
that our solutions are by construction self-similar with $\rho\propto
r^{-3/2}$ and $p\propto r^{-5/2}$.  If $\gamma$ is exactly equal to
$5/3$, then a flow with these radial dependences is automatically
isentropic.  However, for $\gamma\neq5/3$, there is an entropy
gradient in the flow such that the entropy increases inwards whenever
$\gamma<5/3$, i.e. $\epsilon'>0$.  This means that the entropy of each
accreting gas element increases with time at a rate proportional to
$\epsilon'$.  The entropy has to be generated of course by the viscous
energy dissipation, described by the four terms in the right of eq
(2.19).  Usually, the dissipation is dominated by the two terms
proportional to $\Omega^2$ amd $(d\Omega/d\theta)^2$ which arise from
the $r\phi$ and $\phi z$ components of the shear stress.  Thus the
dissipation is proportional to $\Omega^2$.  In order to achieve
self-consistency, the magnitude of $\Omega$ in the self-similar
solution is adjusted such that the dissipated energy exactly matches
the energy that is required to maintain the self-similar entropy
gradient.  Since the latter is proportional to $\epsilon'$ and since
$v\sim -\alpha$, we thus see from (2.19) that we require
$\Omega\sim(\epsilon')^{1/2}$.  This is exactly what we see in our
numerical solutions.

A somewhat interesting feature of eq (2.19) is that even when
$\Omega\rightarrow0$, the right hand side still remains non-zero.
This is because of the terms $v^2$ and $(dv/d\theta)^2$, the first of
which is the viscous dissipation arising purely from the geometric
convergence of the flow due to the spherical geometry, while the
second represents the $rz$ component of the shear stress.  These two
terms imply a certain minimum level of viscous dissipation even in a
very slowly rotating flow.  This dissipation will of course cause the
entropy to increase inwards, and by the arguments given above we see
that $\epsilon'$ has to be greater than a certain minimum value.
Indeed, we have discovered from our numerical experiments that
self-similar solutions exist only for $\epsilon^{\prime}\ \sgreat\
C\alpha^2$, where $C$ is a constant of order unity.  The limit is
exactly of the form we expect from eq (2.19).  Thus for large values
of $\alpha$, the parameter $\gamma$ needs to deviate significantly
from $5/3$ and/or $f$ needs to be quite different from unity in order
to have a self-similar flow.  On a related point, we note that there
is a second branch of solutions which corresponds to non-rotating
purely spherical accretion.  These solutions are quite distinct from
the $\Omega\rightarrow0$ limit of the rotating solutions discussed
here and are in fact closely related to Bondi (1952) spherical
accretion.  We discuss these solutions in Appendix B.

In addition to the solutions described so far, we have found (for
somewhat large values of $\epsilon^{\prime}/\alpha^2$) other solutions
where $\Omega$ reverses sign one or more times as a function of
$\theta$.  These higher-order solutions come in two parities.
Solutions with even parity have an even number of nodes in
$\Omega(\theta)$ between $\theta=0$ and $\theta=\pi$.  These solutions
satisfy the boundary conditions (2.22) at the equator, and have
rotation profiles which are symmetric between the two hemispheres.
The lower right panel of Fig. 2 shows isodensity contours of one such
solution with two nodes where the flow down the two poles rotates in
one sense while the flow in the equatorial plane rotates with the
opposite sense.  Solutions with odd parity have an odd number of
nodes, and the rotation profiles in the two hemispheres are reversed
with respect to each other.  These solutions satisfy the boundary
condition $\Omega=0$ at $\theta=\pi/2$.  We do not expect any of these
higher-order solutions to be relevant except in rare cases where the
initially infalling gas happens to have reversals in the sign of the
angular momentum as a function of $\theta$.  In the rest of the paper
we restrict ourselves to the nodeless fundamental solutions.

The results described so far correspond to the particular form of
viscosity given in eq. (2.6).  To check how sensitive the results are
to the viscosity prescription, we obtained solutions corresponding to
a second law.  By dimensional analysis we see that self-similarity is
possible only if $\nu$ scales with radius as $r^{1/2}$.  Therefore, as
our alternate prescription we used $\nu=\alpha c_sr$.  We found that
the solutions with this viscosity law are very similar to those
described above.  Therefore, none of our results are special to the
particular viscosity prescription we have adopted.

\bigskip
\noindent{\bf 3.2. Comparison with Height-Integrated Solutions}
\bigskip
One of the primary aims of this study was to check the validity of the
height-integrated approximation for advection-dominated flows.  As we
have seen, accretion flows become very nearly spherical when they are
advection-dominated ($f\rightarrow1$).  From this it might appear that
the height-integrated approximation, which is based on a disk-like
picture of the flow, would be particularly inappropriate in this
limit.  We investigate the issue quantitatively.

Spruit et al. (1987) and NY showed that the height-integrated
equations have the following analytical self-similar solution,
$$
\eqalignno{
(v)_h & =-(5+2\epsilon^{\prime}){g(\alpha,\epsilon^{\prime})\over3\alpha}
\approx -{3\alpha\over (5+2\epsilon^{\prime})}, & (3.1) \cr
(\Omega)_h & =\left[{2
\epsilon^{\prime}(5+2\epsilon^{\prime})g(\alpha,\epsilon^{\prime})
\over 9\alpha^2}\right]^{1/2}
\approx \left[{2\epsilon^{\prime}\over
5+2\epsilon^{\prime}}\right]^{1/2}, & (3.2) \cr
(c_s^2)_h & =
{2(5+2\epsilon^{\prime})\over 9}{g(\alpha,\epsilon^{\prime})\over\alpha^2}
\approx {2\over 5+2\epsilon^{\prime}}, & (3.3)
}
$$
where the second relation in each equation refers to the limit
$\alpha\ll1$, and the subscript $h$ is to remind us that these
expressions correspond to the height-integrated approximation.  The
function $g(\alpha, \epsilon^{\prime})$ is given by
$$
g(\alpha,\epsilon^{\prime})\equiv \left[{1+{18\alpha^2\over
(5+2\epsilon^{\prime})^2}}
\right]^{1/2}-1. \eqno (3.4)
$$

We now take the exact solutions calculated in this paper and extract
from them three fiducial values each of $v$, $\Omega$ and $c_s^2$.
First, we consider the values of the variables at the mid-plane,
$\theta=\pi/2$; we refer to these values as $(v)_m$, $(\Omega)_m$ and
$(c_s^2)_m$.  As our second estimate, we compute spherically-averaged
values, e.g.
$$
\langle v\rangle_\theta={\int_0^{\pi/2}v(\theta)\rho(\theta)
d\theta\over \int_0^{\pi/2}\rho(\theta)d\theta},
\eqno(3.5)
$$
with $\langle\Omega\rangle_\theta$ and $\langle c_s^2\rangle_\theta$
defined similarly.  Finally, we compute $z$-averaged values,
which correspond to cylindrical averages parallel to the rotation
axis.  In this case, we define the averages according to
$$
\langle v\rangle_z={\int_0^{\infty}v(z)\sin(\theta)\rho(z)dz\over
\int_0^{\infty}\rho(z)dz},
\eqno(3.6)
$$
$$
\langle\Omega\rangle_z={\int_0^{\infty}\Omega(z)\rho(z)dz\over
\int_0^{\infty}\rho(z)dz},
\eqno(3.7)
$$
$$
\langle c_s^2\rangle_z={\int_0^{\infty}c_s^2(z)\rho(z)dz\over
\int_0^{\infty}\rho(z)dz}.
\eqno(3.8)
$$
Figure 3 compares these three fiducial values of $v$, $\Omega$ and
$c_s^2$ from the exact solutions with the corresponding
height-integrated values, $(v)_h$, $(\Omega)_h$, $(c_s^2)_h$, for a
range of $\epsilon^{\prime}$ extending from
$\epsilon^{\prime}=10^{-2}$ to $10^2$.  The results correspond to
small values of $\alpha\ \sles\ 0.1$; in this regime, $\Omega$ and
$c_s^2$ are independent of $\alpha$ while $v$ is simply proportional
to $\alpha$.

When $\epsilon^{\prime}\gg1$, we are in the cooling-dominated regime
where the flow resembles standard thin disk accretion.  In this limit,
we expect height-integration to be quite accurate, and indeed we do
find that the height-integrated estimates $(v)_h$, $(\Omega)_h$,
$(c_s^2)_h$ agree well with all three estimates obtained from the
exact numerical solutions.  The only discrepancy is between
$(c_s^2)_h$ and $(c_s^2)_m$, but this can be understood.  These
solutions have coronae as discussed in \S 3.1, and naturally
the midplane value of $c_s^2$ is smaller than either the spherical
average or the $z$-average.

When $\epsilon^{\prime}$ falls below unity the flow becomes quite
spherical, and we would expect height-integration to be less valid.
As expected, we find that the $z$-averaged values of $v$, $\Omega$
and $c_s^2$ differ significantly from the height-integrated estimates.
For instance, at $\epsilon^{\prime}=10^{-2}$, we find $\langle v\rangle_z
=0.32(v)_h$, $\langle\Omega\rangle_z =0.42(\Omega)_h$, $\langle
c_s^2\rangle_z =0.46(c_s^2)_h$.  Thus, height-integration leads to
fairly large errors at the level of factors $\sim$ 2--3.

However, Fig. 3 reveals a surprise, viz. that the height-integrated
estimates agree very well with both the midplane values and the
spherically-averaged values of the exact solutions.  For instance, at
$\epsilon^{\prime}=10^{-2}$, we find $(v)_m =1.28(v)_h$, $(\Omega)_m
=1.13(\Omega)_h$, $(c_s^2)_m =1.00(c_s^2)_h$ for the midplane values,
and $\langle v\rangle_\theta =0.82(v)_h$, $\langle\Omega\rangle_\theta
=1.09(\Omega)_h$, $\langle c_s^2\rangle_\theta =1.00(c_s^2)_h$ for the
spherical averages.  The spherical averages show particularly good
agreement with the height-integrated estimates over the entire range
of $\epsilon^{\prime}$ from 0 to infinity, with no error exceeding
$20\%$.

This result suggests that one should interpret the height-integrated
equations, not as averages over cylindrical height $z$, but rather as
averages over spherical polar angle $\theta$ at a fixed $r$.  Once
this is done, height-integration is a good approximation even for
nearly spherical flows.  The height-integrated equations, especially
in the form of the so-called ``slim disk'' equations (Abramowicz et
al. 1988), have become popular in recent years for modeling the
dynamics of accretion flows.  Until now, it has not been clear exactly
how slim a disk has to be in order for the equations to be valid, and
also exactly what kind of an average the solutions represent.  Based
on the results presented here, we suggest that the slim disk equations
may be applied virtually to any accretion flow around a point mass,
however non-slim the flow may be, and that the results should be
interpreted as spherical averages rather than as $z$-averages.  We
caution, however, that this suggestion is based on the properties of a very
special class of self-similar solutions, and needs to be tested on
non-self-similar flows.

\bigskip
\noindent{\bf 3.3. The Bernoulli Parameter}
\bigskip

The Bernoulli parameter $Be$, defined as the sum of the kinetic
energy, the potential energy and the enthalpy of the accreting gas, is
of interest in accretion flows because it measures the likelihood that
outflows or winds may originate spontaneously (NY).  An adiabatic flow
has a constant $Be$ along streamlines.  If $Be$ is positive for any of
the accreting gas, then this gas can potentially reach infinity with a
net positive kinetic energy.

Let us normalize $Be$ in our self-similar solutions by the square of
the local free-fall speed, $v_{ff}=\Omega_Kr$, and consider the
dimensionless parameter
$$
b\equiv {Be\over \Omega_{K}^2r^2}={1\over 2}v^2+{1\over 2}
(\Omega\sin\theta)^2-1+{\gamma\over\gamma-1}c_s^2.
\eqno(3.9)
$$
The parameter $b$ is a function of $\theta$ but independent of $r$.
In NY we showed that the Bernoulli parameter is positive in
height-integrated advection-dominated flows, and suggested that this
may explain the frequent occurrence of outflows and winds in many
accretion systems.  However, because the result was obtained through
the height-integrated equations, it had to be treated with caution.
We now consider the behavior of $b$ in the exact self-similar
solutions obtained in this paper.

Fig. 4 shows $b(\theta)$ for a sequence of solutions.  All the
solutions have the same $\epsilon=0.333$ ($\gamma=1.5$), but the
advection-domination parameter $f$ varies over the range $f=1, ~0.446,
{}~0.33, ~0.033, ~0.0033, ~0.0011$, which means that
$\epsilon^{\prime}=0.33, ~0.75, ~1, ~10, ~100, ~300$.  We confirm the
basic result of NY that advection-dominated flows have positive $b$.
Furthermore, we see that for $f>0.446$, $b(\theta)>0$ at all $\theta$.
This means that for flows that are so highly advection-dominated, the
{\it entire} gas has positive Bernoulli parameter.  Even in the limit
of extreme cooling ($f\rightarrow0$), there is always some gas near
the rotation axis ($\theta\rightarrow0$) which has a positive $b$.
The reason for this is the corona which we mentioned in \S3.1.  Even
though the disk cools efficiently and is geometrically thin, it still
has a hot virial corona above it and this region of the flow can
acquire a positive $b$ especially close to the rotation axis.

These results make the connection between accretion and outflows much
stronger than we suspected based only on the height-integrated work of
NY.  A case can be made now that perhaps all accretion flows, whether
advection-dominated or cooling-dominated, are capable of producing
outflows.  The main difference between the two kinds of flow may be
only a quantitative one, viz. the former probably produce more
powerful outflows than the latter.  Also, it would appear that
outflows will prefer to form along the rotation axis, since this is
where $b$ is most positive in all cases.  A bipolar morphology is thus
natural.  The argument is made even stronger by the results on
convection discussed in the next section.

It is important to emphasize that the positivity of $b$ does not imply
a violation of energy conservation.  In viscous accretion flows, the
energy content of a parcel of gas is modified by viscous transport of
energy from one radius to another and from one $\theta$ to another.
Let us consider only the radial flux for simplicity.  The radial flux
of energy transported by viscosity is equal to the quantity (the
radial angular momentum flux) $\times$ (the angular velocity)
$\sim\dot M v_\phi^2$.  This energy flux is directed outward and has a
negative divergence.  Therefore, it deposits energy at each radius.
For a given $\dot M$, the rate of deposition of energy is maximum in
the case of a standard thin accretion disk, where $v_\phi=\Omega_Kr$.
However, because these disks also cool efficiently, the deposited
energy does not produce an enhancement of $b$ (in fact, $b=-0.5$), but
rather leads to an enhancement of the locally radiated flux.  This
explains the well-known result that the radiative flux emerging from a
thin disk (except near the inner edge) is three times larger than the
rate at which gravitational energy is released locally (e.g. Frank,
King \& Raine 1989).  As the parameter $f$ increases and the flow
becomes more advection-dominated, the viscous energy flux actually
decreases in magnitude because $v_\phi$ becomes smaller than
$\Omega_Kr$.  However, because the cooling is also less efficient, a
larger fraction of the energy is retained by the gas and therefore,
paradoxically, $b$ actually becomes larger.  As we have seen, when $f$
exceeds a critical value, $b$ can actually become positive over the
entire flow.

The $\theta$ component of the viscous stress causes angular
redistribution of the energy at a given $r$.  The direction of this
flux is such as to enhance $b$ at the poles relative to the equator.
The result is that the pole always has a positive $b$ even in the
limit of small $f$.

The self-similar solution is very special in that it has an infinite
source of energy at $r=0$ which can be transported outward by
viscosity.  This explains how the entire solution can have a positive
$b$.  If we consider a non-infinite flow which is terminated at a
finite inner radius $r_i$, then for radii close to $r_i$ the viscous
energy flux will be less than the self-similar value.  The gas near
the inner edge will therefore have negative $b$.  However, once we are
reasonably far from the inner edge (say $r> {\rm few}\times r_i$), the
self-similar value of $b$ will be achieved and beyond this radius the
flow will be indistinguishable from the self-similar form.  Overall,
the deficit of $b$ near the inner edge will compensate for the
positive $b$ elsewhere, ensuring that energy is conserved globally.

\bigskip
\noindent{\bf 3.4. Convection}
\bigskip

In NY, we showed that advection-dominated flows have entropy
increasing inwards and therefore that these flows are intrinsically
unstable to convective instabilities.  This point was made earlier by
Begelman \& Meier (1982).  We discuss here the role of convection in
the solutions described in this paper.

The left-hand side of eq (2.19) is proportional to $vTds/dr$, where
$ds/dr$ is the radial derivative of the entropy.  Since $v$ is
negative and the right-hand side of eq (2.19) is positive (it consists
only of dissipation terms), we see that $ds/dr$ is negative at all
$\theta$, and therefore that the flow is convectively unstable at all
angles.  In any simple theory of convection we expect the convective
flux to be parallel to the local pressure gradient since this is the
direction associated with the buoyancy force.  (In principle, the flux
could be in a different direction in a rotating flow because of
anisotropic transport, but we ignore this complication.)  We have seen
earlier that whenever the flow is advection-dominated ($f\rightarrow1,
{}~\epsilon ^{\prime}\ \sles\ 1$) the flow is nearly spherical and the
pressure gradient is almost radial.  Therefore, we expect convection
to act primarily in the radial direction.  This is very different from
the vertical convection which is usually discussed in the context of
thin accretion disks (e.g. Ryu \& Goodman 1992).

When there is convection we can expect it to have a back reaction
on the basic flow.  To estimate the magnitude of
the effect, let us rewrite the energy equation (2.19)
with the convective contribution included:
$$
-{3\epsilon\over 2}{\rho v_rc_s^2\over r}=-{\bf\nabla}\cdot
{\bf F_c}+{f\alpha\rho c_s^2\over\Omega_K}
\left[{3v_r^2\over r^2}+{1\over r^2}\left({dv_r\over d\theta}\right)^2
+{9\over4}\sin^2\theta\Omega^2
+\sin^2\theta\left({d\Omega\over d\theta}\right)^2\right].\eqno (3.10)
$$
Here ${\bf F_c}$ is the convective energy flux, and $\rho, ~v_r,
{}~\Omega, ~c_s$ refer to the physical variables and not the scaled
functions defined in eqs (2.11)--(2.15).  The term on the left of eq
(3.10) is the divergence of the advected entropy flux, $\rho T{\bf
v}\cdot{\bf\nabla}s$, which we refer to as the ``advection term.''
The first term on the right is the net deposition of energy due to the
inflow of convective energy flux into the gas --- the ``convection
term'' --- and the second term is the energy deposition due to viscous
dissipation (reduced by our usual factor $f$).

To estimate the magnitude of the convection term, we
note that we expect the convective flux to be proportional to the
entropy gradient with some effective diffusion constant $K_c$.
Further, we may write $K_c$ approximately in a form similar to the
$\alpha$-prescription for viscosity.  Thus, we have
$$
{\bf F_c}\approx -K_c\rho T({\bf\nabla}s\cdot{\hat r}){\hat r}
\approx-{\alpha_cc_s^2\over\Omega_K}\rho T({\bf\nabla}s\cdot{\hat r})
{\hat r}, \eqno (3.11)
$$
where we have assumed that ${\bf F_c}$ is parallel to the unit radial
vecotr $\hat r$ because the pressure gradient is primarily radial.
Since ${\bf\nabla}s\cdot{\hat r}$ is negative we see that the
convective flux flows outwards.  In an advection-dominated
self-similar flow $c_s^2\ \sles\ 2\Omega_K^2r^2/5$ (see eq 3.3), and
$F_c$ is proportional to $r^{-3}$.  Therefore, the convection term
becomes
$$
-{\bf\nabla}\cdot{\bf F_c}={F_c\over r}\ \sles\
{3\over5}\alpha_c\epsilon\rho c_s^2\Omega_K.\eqno (3.12)
$$
We see that this term has the same form as the advection term on the
left of equation (3.10).  We can therefore compare the two terms
directly.  Figure 5 shows the ratio of the convection and advection
terms for one of our advection-dominated solutions, assuming
$\alpha_c=\alpha/2$.  The shape of the curve may be understood as
follows.  For small $\epsilon^{\prime}$, the radial velocity $v$ is
given quite accurately by
$$
v(\theta)\approx -{3\over4}\alpha\sin^2\theta\Omega_Kr.\eqno (3.14)
$$
Therefore, we expect
$$
{{\rm Convection}\over{\rm Advection}}\approx
{8\over15\sin^2\theta}{\alpha_c\over\alpha}.\eqno (3.15)
$$
The ratio is smallest in the equatorial plane ($\theta\rightarrow\pi/2$)
and diverges towards the rotation axis ($\theta\rightarrow0,\pi$).

On general grounds, we would argue that $\alpha_c$ cannot exceed
$\alpha$.  The reason is that $\alpha$, which describes the viscous
transport of angular momentum, is expected to have contributions from
many sources of viscosity such as magnetic stresses, convection, and
other fluid instabilities.  Convective energy transport on the other
hand depends only on the strength of the convection.  We therefore
expect $0\leq\alpha_c\leq\alpha$, with the upper equality being
achieved only in the limit of massively efficient convection with no
source of viscous stress other than convection.  In fact, if magnetic
stresses always saturate at equipartition (e.g. Gammie, Hawley \&
Balbus 1994), then the inequality becomes $0\leq\alpha_c\leq\alpha/2$.

Figure 5 and eq (3.15) show that for a wide range of $\theta$, the
convection term is smaller than the advection term, so that convection
has only a moderate effect on the flow.  To see this in more detail,
we take the convection term to the left-hand side of eq (3.10) and
combine it with the advection term (NY).  This gives
$$
{\rm Advection-Convection}={3\epsilon\over2}{\rho
vc_s^2\over r} \left(1-{8\over15\sin^2\theta}
{\alpha_c\over\alpha}\right).\eqno (3.16)
$$
With this modification, eq (2.19) continues to be valid
except that we have to rewrite $\epsilon^{\prime}$ as
$$
\epsilon^{\prime}={\epsilon\over f}
\left(1-{8\over15\sin^2\theta}{\alpha_c\over\alpha}\right).\eqno
(3.17)
$$
The new factor in parentheses modifies the value of
$\epsilon^{\prime}$.  So long as this factor is positive, it has only
a minor effect on the solution.  The value of $\epsilon^ {\prime}$ is
reduced (in an angle-dependent way because of the $\sin^2\theta$) and
the flow becomes effectively more advection-dominated.  The overall
structure of the flow, however, remains basically unchanged.  Since
the convection term is smaller than the advection term, the timescale
on which convection changes the entropy profile is longer than the
advection timescale, and there is just not enough time for convection
to have a large effect.

There is however a critical angle $\theta_{crit}$,
whose value is given by eq (3.16),
$$
\theta_{crit}\approx\sin^{-1}\left[\left({8\alpha_c\over
15\alpha}\right)^{1/2}\right],\eqno (3.18)
$$
below which convection dominates over advection.  This means that
for $\theta<\theta_{crit}$ the flow {\it is} strongly modified.
Indeed, at these angles, convection will overwhelm the advection and
the whole self-similar flow will break down.  At these angles, we
expect that convection will rapidly transfer entropy outwards.  Since
the Bernoulli constant is positive at these angles (see
\S3.3) the gas at large radius will acquire a great deal of
positive energy and in all likelihood will flow out supersonically.

We thus have a plausible scenario for the formation of bipolar
outflows.  Our qualitative picture is that the accretion occurs
primarily in a thick equatorial belt with
$\theta_{crit}<\theta<\pi-\theta_{crit}$.  Over this zone the solution
is almost of the self-similar form described in this paper, and
although it is convective the convection has only a minor effect on
the parameters of the flow.  However, for $\theta<\theta_{crit}$ and
$\theta>\pi-\theta_{crit}$, the convection becomes strong enough to
transfer energy outward and to disrupt the self-similar accretion.
This gas has a positive Bernoulli constant (\S3.3), and therefore very
likely will be driven out in an outflow.  Of course, we still don't
have a self-consistent description of the outflow region, since all we
have shown is that our self-similar advection-dominated solution is
violently unstable in the region around the rotation axis.
Nevertheless, we feel that our picture is plausible and that some
outflows at least may well be generated in this manner.

Figure 5 shows that for $\alpha_c=\alpha/2$, we have
$\theta_{crit}\sim 33^o$.  The outflow region is therefore not very
wide.  Moreover, the amount of mass involved in the outflow region is
quite small.  Let us assume that the mass outflow rate $\dot M_{out}$
is equal to the mass accretion rate in the original self-similar
solution for $\theta<\theta_{crit}$.  While the density $\rho$ is
essentially independent of $\theta$ in an advection-dominated flow,
the accretion velocity varies as $v\propto\sin^2\theta$ (eq 3.14) and
is very small in the region around the axis.  From our numerical
solution we calculate that $\dot M_{out}\sim 0.017\dot M$ for
$\alpha_c=\alpha/2$.  Therefore, only a small fraction of the
accreting mass participates in the outflow.  Indeed this estimate of
$\dot{M_{out}}$ is probably too high since it assumes that the
original self-similar flow has been set up in the polar regions and is
then turned round into an outflow.  In practice we imagine that the
outflow will clear out two conical regions and that the mass for the
outflow will be supplied by pressure gradients in the upper layers of
the equatorial inflow.  The mass outflow rate will then be even
smaller than our estimate.

In any case, the important point is that, in the very nature of the
flow, the mass in the outflow region acquires the most positive energy
(or Bernoulli constant) at the expense of the rest of the accreting
material.  It is therefore ejected with a speed comparable to the free
fall speed at the radius from which the outflow originates.  The most
energetic ejected material will have a speed at infinity comparable to
the virial speed at the surface of the accreting star.

\bigskip\bigskip
\noindent{\bf 4. Summary and Discussion}
\bigskip

The main aim of this investigation was to obtain axisymmetric
self-similar advection-dominated flow solutions in three dimensions
without using the height-averaging approximation.  We have succeeded
in this enterprise.  The solutions we have presented are obtained by
numerically solving a general sixth-order system of differential
equations, where the only serious approximation we have made is the
use of an isotropic $\alpha$ viscosity.  To our knowledge, these
solutions are one of the very few fully self-consistent, axisymmetric,
rotating steady state flows known with non-trivial viscous
interactions.  Related previous work has been published by Begelman \&
Meier (1982), Liang (1988) and Henriksen \& Valls-Gabaud (1994).
However, some of these other solutions have unphysical boundaries
where the assumptions break down, whereas our solutions describe fully
consistent equilibrium flows which fill the entire $r\theta$ plane.

Our solutions span a two-parameter family labeled by the viscosity
parameter $\alpha$ (see eq. 2.6 for the definition) and a
thermodynamic parameter $\epsilon^{\prime}$ (eq. 2.20).  The latter is
a function of the ratio of specific heats $\gamma$ of the accreting
gas and the fraction $f$ of the dissipated energy which is advected
with the gas.  For a given $\alpha$, solutions with large values of
$\epsilon^{\prime}$ behave like standard thin disks, as might be
expected since these solutions correspond to $f\rightarrow0$ and so
advect very little energy.  In the opposite {\it advection-dominated}
limit, which corresponds to $f\rightarrow1$ or $\epsilon^{\prime}<1$,
our solutions describe nearly spherical flows which rotate at much
below the Keplerian rate.  These advection-dominated solutions have
very similar properties to the approximate solutions derived in NY.

We emphasize that the solutions we find are not the usual tori with
steep funnels that are normally considered when ``thick accretion
disks'' are discussed (e.g. Begelman \& Meier 1982, Frank et
al. 1992).  Our solutions are much more like slowly rotating stars,
except that they are not static but involve viscously-driven settling
flows.  The question now arises: do our equations permit solutions
with empty funnels, as described by Begelman \& Meier (1982)?  As we
discuss in the Appendix, this must be considered an open question at
this time.

We have compared the numerical three-dimensional solutions of this
paper with the approximate two-dimensional height-integrated solutions
described by Spruit et al. (1987) and NY.  Because some of our
solutions are nearly spherical, we might expect the height-integrated
solutions to be in error by quite large factors.  Instead, we find
that the height-integrated solutions agree very well (errors $<20\%$)
with spherically-averaged quantities from the exact numerical
solutions.  This is very encouraging since it means that simple
height-integrated equations such as the ``slim disk'' equations
(Abramowicz et al. 1988) may be valid over a much wider range of
conditions than suspected before.  The result also clarifies how to
interpret the height-integrated equations.  One should consider these
equations to represent the properties of {\it spherically-averaged}
quantities in the accretion flow (averages over spherical polar angle
$\theta$ at fixed spherical radius $r$) rather than of vertically
averaged quantities (averages over vertical height $z$ at fixed
cylindrical radius $R$).  This interpretation is, however, based on a
very special class of solutions and should be checked on less special
flows.

Advection-dominated accretion flows have a very unique feature in that
they are characterized by positive values of the Bernoulli parameter
$b$ (eq 3.9).  This result was discovered by NY using the
height-integrated equations, and is confirmed here with more exact
calculations.  We interpret a positive Bernoulli parameter as an
indication that the accreting gas may be able spontaneously to
generate winds and outflows.  In some of our highly
advection-dominated solutions we find that the entire accreting gas
has a positive $b$ (e.g. the solutions with $f\geq0.446$ in Fig. 4).
These flows may be particularly susceptible to violent outflows.
Rather surprisingly, even solutions with efficient cooling
($f\rightarrow0$) have hot low density coronae with positive $b$.
This might mean that even regular thin accretion disks may always be
accompanied by advection-dominated coronae which can drive low-density
winds.  Further, both in the advection-dominated and cooling-dominated
limits, $b$ is maximally positive along the rotation axis.  This
suggests that whenever there is an outflow, it is likely to have a
generic bipolar morphology.

Another interesting property of advection-dominated flows is that they
are always convectively unstable, as was noted originally by Begelman
\& Meier (1982).  Because advection-dominated flows have almost
spherical isobars, we argue that the convection will occur almost
purely in the radial direction.  This is very different from the
``vertical convection'' which has been studied in the context of thin
accretion disks (e.g. Ryu \& Goodman 1992).

Convection in advection-dominated flows may be a source of turbulent
viscosity.  Although viscosity is required in order to have any
accretion at all, for long it was unclear what the source of the
viscosity in accretion disks may be since no linear hydrodynamic
instability could be identified in Keplerian disks.  With the recent
work of Balbus \& Hawley (1991), interest therefore shifted to
magnetic stresses generated through an MHD instability.  The only
model that we are aware of that explains accretion disk viscosity
through a purely hydrodynamic model is that of Dubrulle (1992) who
uses a finite amplitude instability. This model, however, results in a
very small $\alpha\ll 1$.  The convective instability which we
identify in our advection-dominated flows is a true linear
instability.  Furthermore, if the flow is self-similar, then the
convection also will automatically be self-similar, and this means
that the convective viscosity will necessarily have a form similar to
the $\alpha$-scaling given in eq. (2.6).  Our models are thus rather
close to being perfectly self-consistent in the sense that the
viscosity is not imposed in an ad hoc manner but could, if necessary,
itself be determined self-consistently from an instability in the
flow.

By comparing the convective time scale with the advective time scale,
we find that for most regions of our solutions, convection acts only
as a moderate perturbation which changes the parameters of the flow by
a modest amount but does not destroy the overall structure of the
flow.  However, for a range of angles around the rotation axis,
$\theta<\theta_{crit}\sim 30^o$, the convection is capable of
overwhelming the advection and may transport entropy outwards more
rapidly than it can be carried in by accretion.  Since this region of
the flow is also the zone with the largest Bernoulli parameter, we
suggest that the violent convection and the positive $b$ will together
act to reverse the local accretion flow into a bipolar outflow.  The
speed of the ejected material at infinity will be of order the
free-fall speed at the radius of origination of the material.  We
imagine that the outflow will be fed by surface material from the
equatorial inflowing gas.

We feel that this scenario is a plausible and generic mechanism to
produce outflows.  For the mechanism to work, we require the
outflowing material to remain adiabatic and to retain its Bernoulli
parameter long enough to be accelerated beyond the escape velocity.
Given this requirement, we suggest that an outflow can be created
merely with viscous and convective redistribution of energy in the
manner we have described, without any additional agencies.
Hydrodynamic models of outflows have been discussed before by Eggum,
Coroniti \& Katz (1988), Liang (1988) and Henriksen
\& Valls-Gabaud (1994).  In the former two papers, the outflow is
accelerated by radiation pressure, while magnetic stresses appear to
play a role in the last work.  Radiation and magnetic fields doubtless
help, but we suggest that outflows are possible even without them.

Note that we do not at this time have a truly self-consistent
description of the outflow.  By insisting on self-similarity and a
constant $\dot M$ at all radii, we are forced to set $v_\theta=0$, and
because of this our solutions do not have the option of diverting any
of the accreting mass into an outflow.  The condition of constant
$\dot M$ is relaxed by some other authors (e.g. Liang 1988, Henriksen
\& Valls-Gabaud 1994), and they do find solutions with outflows.  It
is not clear, however, whether their outflows are driven by the same
mechanism that we suggest.  It would be of interest to reformulate our
problem with a non-constant $\dot M$ to investigate whether or not
self-similar outflows are possible with these equations.

The principal characteristic of an advection-dominated accretion flow
is that most of the dissipated energy is stored within the gas.  This
can happen at very low accretion rates when the gas density is so low
that cooling through bremsstrahlung is inhibited.  Such an effect was
seen in detailed numerical models of boundary layers in cataclysmic
variables (Narayan \& Popham 1993).  Alternatively,
advection-domination may occur when the accretion rate is extremely
high.  In this case, the optical depth of the flow becomes so large
that the radiation is unable to escape in less than a flow time.
Popham et al. (1993) found that this may happen in FU Orionis systems.
Advection-domination has been considered also in disks around
supermassive black holes.  Rees et al. (1982) discussed ion tori at
very low accretion rates, while Begelman (1978) discussed radiation
trapping at very high $\dot M$ where the photon diffusion time scale
becomes longer than the radial inflow time scale (or advection time
scale).

If advection-dominated conditions are present over a sufficiently wide
range of radius in an accreting source, then we may expect the flow to
approximate the self-similar solutions discussed here (cf, Fig. 1 in
NY).  The various properties of these solutions then become relevant.
To summarize:

\noindent
1. The flow is definitely not disk-like in morphology.  In fact, the
closest analog to our solutions in the accretion literature is Bondi
(1952) spherical accretion.  However, our flows differ in important
ways from the Bondi problem.  The gas in our solutions does rotate, it
has non-trivial viscous interactions through which angular momentum is
transported outward, and it has energy dissipation just as in ordinary
accretion disks.

\noindent
2. The angular velocity is significantly sub-Keplerian and this may
have important implications for the spin-up of accreting stars.  Stars
which spin up through an advection-dominated mode of accretion are
likely to reach a steady state with a rotation rate much below the
``break-up limit.''  This may be one solution to the angular momentum
problem which is discussed frequently in the context of star
formation.

\noindent
3. The radial accretion velocity is typically high in
advection-dominated flows.  Roughly, the velocity scales as
$v\sim\alpha c_s$.  In a cooling-dominated disk, $c_s\ll v_{ff}$, where
$v_{ff}$ is the free-fall velocity, and so $v\ll v_{ff}$.  However,
advection-dominated flows have $c_s\sim v_{ff}$ and so for a
reasonable $\alpha\sim0.1$, we find $v\sim 0.1v_{ff}$.

\noindent
4. The Bernoulli parameter is positive over most of the flow (except
the regions very close to the inner edge), and there is also likely to
be violent convection especially close to the rotation axis.  A fairly
substantial bipolar outflow is quite likely under these conditions
provided only that the material can remain adiabatic during the
acceleration phase of the outflow.

\noindent
5. The convective motions will transport both energy and angular
momentum and will be a source of viscosity.  This has the advantage
that, even if for some reason the Balbus \& Hawley (1991) instability
were to be ineffective, convective viscosity can still keep the
accretion going.

\noindent
6. By definition, advection-dominated accretion systems are
under-luminous relative to the mass accretion rate.  This is because
the energy is carried along with the accreting gas as heat instead of
being radiated.  If the accreting object is a regular star, this
energy must finally be radiated from the star and will be seen as
stellar emission.  However, if the accretor is a black hole, then most
of the accretion energy can disappear through the horizon.  It would
be very misleading to estimate the mass accretion rate of such a
system from the observed luminosity.

\noindent
7. The spectrum of an advection-dominated flow is likely to be quite
different from that of a cooling-dominated accretion disk.  If the
flow is in the low $\dot M$ optically thin limit, then the temperature
of the emitted radiation will be close to virial and the spectrum will
be unusually hard.  This appears to be the situation in cataclysmic
variable boundary layers for $\dot M\ \sles\ 10^{-9.5}M_\odot{\rm
yr}^{-1}$ (Narayan \& Popham 1993).  In the opposite radiation-trapped
limit at high $\dot M$, the photosphere will be much farther out than
usual and the spectrum will be unusually soft.  This is apparently the
case in FU Orionis systems (Popham et al. 1993).  Similar effects must
be present in accreting neutron star and black hole systems, but this
remains to be investigated.

We conclude with a final speculative comment on accretion disk
coronae.  An interesting feature of our solutions in the limit of
large $\epsilon^{\prime}$ (efficient cooling) is that they have low
density hot gas on top of an equatorial thin disk (see the
$\epsilon^{\prime}=10$ solution in Fig. 1).  The hot gas resembles the
ad hoc coronae that various researchers have invoked in models of
accretion disks.  We speculate (i) that coronae are a natural and
inevitable feature of any thin disk, (ii) that such coronae are best
described as advection-dominated flows rather than as static
atmospheres, and (iii) that these coronae will themselves have
outflows for the reasons discussed in this paper, though with much
smaller mass loss rates than in fully advection-dominated flows.
Whenever a disk plus corona structure is formed, the accreting
material must be divided in some self-consistent manner between the
thin cool disk and the thick hot corona.  One possibility is that the
corona always has just enough mass in it to be {\it marginally}
advection-dominated, i.e. to have $f_{corona}\sim 1/2$.  Detailed
calculations with radiative transfer are needed to confirm whether
such a model will be self-consistent.

\bigskip\bigskip\noindent
{\it Acknowledgements}: We thank Mitch Begelman and Henk Spruit for
useful comments.  This work was supported in part by the NSF under
grant AST-9148279.

\vfill\eject
\noindent{\bf APPENDIX A: Comparison with Begelman and Meier (1982)}
\bigskip

Begelman \& Meier (1982, hereafter BM) considered a problem very
similar to the one described in this paper, and it is useful to
compare the two approaches.  BM restricted their attention to a
radiation-dominated accretion flow which is fully advection-dominated.
In the language of this paper, their model corresponds to the particular
case, $\gamma = 4/3$, $\epsilon^{\prime} = \epsilon = 1$.  They
considered however a more general viscosity model than we did, viz.
$$
\nu \propto p^{\alpha'}\rho^{\beta'}r^{\gamma'}g(\theta),\eqno(A1)
$$
where $\alpha',~\beta',~\gamma'$ are three arbitrary exponents.
$g(\theta)$
is a general function of $\theta$, though in actual practice, BM set
$g(\theta)$ =
constant in their calculations.  By imposing additional requirements like
self-similarity, BM restricted their viscosity model to
satisfy
$$
\beta' = -\alpha',\qquad 1 + \alpha' - \gamma'= 1/2, \eqno(A2)
$$
so that they had effectively a one-parameter family of viscosity models
parametrized by $\alpha'$.
Our model of viscosity, eq. (2.6), corresponds to the
particular choice,
$$
\alpha' = 1,\qquad \beta' = -1,\qquad \gamma' = 3/2, \eqno(A3)
$$
and is one member of this family.  We have also carried out a few
calculations with another model corresponding to $\alpha' = 1/2,
{}~\beta' = -1/2,~\gamma' = 1$.

Starting with the basic equations (2.7) -- (2.10) which we have written
in sec. 2, BM assumed that
$$
v_r, ~ v_{\theta} \ll v_{\phi}, \eqno(A4)
$$
and derived the following second-order differential equation for
$\Omega (\theta)$,
$$
(1-\sin^2\theta \Omega^2)\left(\Omega'' + 3\cot \theta \Omega^{\prime} -
{3\over 4} \Omega\right)={\mu \sin^2\theta\Omega\over 2(\mu +4)}
\left({9\over 4} \Omega^2 + \Omega^{12}\right)
$$
$$
\qquad\qquad
-2(\beta' + 1)\sin^2\theta \Omega'^{2} - (\gamma' + 3\beta' + 4)
\cos \theta \sin\theta\Omega^2\Omega^{\prime}.\eqno(A5)
$$
The parameter $\mu$ is the radial exponent of the pressure, $p\propto
r^\mu$.  For self-similar solutions, $\mu=-5/2$.  Notice that eq. (A5)
is only a {\it second-order} differential equation, whereas the system
of equations we derived in this paper is {\it sixth-order}.  The
simplification arises because of the assumption $v_r \ll v_{\phi}$
(eq. A4).

It is straightforward to derive an equation similar to (A5) starting
with our equations (2.16) -- (2.19).  As discussed in the paper, our
solutions have $v \sim \alpha$.  Therefore the condition $v \ll v_{\phi}$ is
equivalent to the assumption $\alpha \ll 1$.  In fact, all we
need is $\alpha^2\ll1$.  Under this condition, we
can neglect all terms involving $v^2$ and $\alpha v$ in eqs. (2.16), (2.17)
and (2.19).  Eq. (2.16) then gives
$$
c^2_s = {2\over 5} (1-\sin^2\theta\Omega^2), \eqno(A6)
$$
while (2.17) gives
$$
{d\over d\theta} (\rho c_s^2) = \cos \theta \sin \theta \rho\Omega^2.
\eqno(A7)
$$
Eliminating $v$ between (2.18) and (2.19) and substituting (A6) and (A7)
we then find
$$
(1-\sin^2\theta\Omega^2)(\Omega'' + 3\cot \theta \Omega^{\prime} -
{3\over 4} \Omega )= - {5 \over 6\epsilon^{\prime}} \sin^2\theta
\Omega \left( {9\over 4}\Omega^2 + \Omega'^{2}\right)
$$
$$
\qquad\qquad\qquad
-{5 \over 2} \cos \theta \sin \theta \Omega^2 \Omega^{\prime}.\eqno(A8)
$$
This is exactly equivalent to eq. (A5) provided we set $\epsilon^{\prime}
= 1$ (radiation-dominated, full advection) and $\beta' = -1,~
\gamma' = 3/2$ (eq A3), $\mu=-5/2$.  We thus find perfect agreement
between our equations and those of BM so long as $\alpha^2,~ v^2 \ll
1$.

All the solutions we have presented in this paper have $\alpha = 0.1$
or smaller, so that $\alpha^2 \ll 1$.  Furthermore, all of our
solutions have well-behaved non-singular $v(\theta)$ and satisfy $v^2
\ll 1$ at all $\theta$.  Therefore, although our solutions were
obtained by solving the exact sixth-order equations (2.16) -- (2.19),
they do in fact satisfy BM's second-order equation (A5) very
accurately.  These solutions seem to have been missed by BM.

The solutions that BM described in their paper correspond to
flows which extend from the equatorial plane at $\theta = \pi/2$ to a
free surface at $\theta = \theta_F$, inside of which is an empty
funnel.  A somewhat disturbing feature of their solutions is that the
radial velocity $v$ diverges at $\theta = \theta_F$.  The divergence
arises because the ram pressure term $v^{2}/2$ and the poloidal
viscous term $(d/d\theta)(\alpha dv/d\theta)$ in eq. (2.16), which
would normally control the divergence, were eliminated through the
assumption $v_r \ll v_\phi$.  A divergent solution is of course not
compatible with the original assumption $v\ll 1$, and this implies an
inconsistency in the solutions.  Another odd feature of the BM
solutions is that for fixed values of the parameters ($\epsilon',~
\alpha,~\alpha',~ \beta',~\gamma', ~\mu$),
they find a continuous family of self-similar solutions, with a
continuously tunable funnel opening angle $\theta_F$.  We find this
infinity of solutions somewhat disturbing and worry that perhaps a
boundary condition at the free surface may have been missed.  We note
that in our calculations, when we fix $\alpha$ and $\epsilon'$, we
find a unique fundamental solution.  (We do have the curious
higher-order solutions described in \S 3.1, where $\Omega(\theta)$ has
one or more nodes, but these solutions still form only a discrete set,
not the continuous family that BM find.)

We feel that it is important to search for solutions with empty
funnels and free surfaces using the full sixth-order set of equations
(2.16) -- (2.19).  The numerical methods which we have employed in the
calculations described in this paper are incapable of finding such
solutions.  It is therefore a completely open question whether or not
such solutions exist.  Tori and empty funnels have been much discussed
in accretion astrophysics, expecially in the context of AGN (see
Frank, King \& Raine 1992 for a review).  However, except for the work
of BM, there have been no self-consistent dynamical models of such
flows which include viscosity fully.  In our opinion this is an
important problem.

\vfill\eject
\noindent{\bf APPENDIX B: Non-Rotating Spherical Inflow}
\bigskip

In addition to the rotating solutions which are the main focus of the
paper, the self-similar equations (2.16)--(2.19) allow a second
non-rotating spherically symmetric branch of solutions which is
closely related to the Bondi (1952) problem.  To derive this solution
we set $\Omega=0$ and $d/d\theta=0$.  Equations (2.17) and (2.18) then
immediately drop out.  Equation (2.19) gives
$$
-{3\epsilon' v\over2\alpha}=3v^2,\eqno (B1)
$$
which has two solutions for $v$.  One of these is $v=0$, which is the
condition satisfied at $\theta=0$ by all the rotating solutions
described in the paper.  For spherical inflow, we consider the second
possibility, namely
$$
v=-{\epsilon'\over2\alpha}.\eqno (B2)
$$
Substituting this in eq (2.16), we then solve for the sound speed, to
obtain
$$
c_s^2={2-\epsilon'^2/4\alpha^2\over5-\epsilon'}.\eqno (B3)
$$
Equations (B2), (B3) coupled with the self-similar scalings completely
describe this branch of solutions.  We see that these solutions are
allowed whenever $|\epsilon'|<2\sqrt{2}\alpha$ (or $\epsilon'<5$ if
$\alpha$ is large).  Thus, a self-similar form of spherical accretion
is possible so long as $|\gamma-5/3|\ \sles\ \alpha$.

Note that in pure Bondi (1952) spherical flow, a self-similar form of
accretion or outflow is allowed only for a single value of $\gamma$,
viz. $\gamma=5/3$.  In contrast, we find that self-similarity is
possible in the present problem for a range of values of $\gamma$.
The additional freedom arises because we have viscosity.  Even though
the flow is not rotating, it still has a velocity divergence and this
gives rise to the viscous dissipation term $3v^2$ in the right of eq
(B1).  When $\gamma\neq5/3$, self-similarity requires a radial
gradient in the entropy (as discussed in \S3.1).  The solution feeds
this gradient by tuning the magnitude of $v$ so as to supply the
required energy input through viscous dissipation.

The other interesting feature is that we can have either self-similar
spherical inflow or outflow, depending on the sign of $\epsilon'$ (eq
B2).  Since viscosity is always a source of energy, the entropy has to
increase in the direction of the flow.  For $\gamma<5/3$, the entropy
increases inward in the self-similar solution, and so the motion is
also inwards, i.e. we have accretion.  However, for $\gamma>5/3$, the
solution corresponds to entropy increasing outward.  In this case,
therefore, our solution corresponds to a spherical self-similar wind.

Note that the spherical solutions described here may be either
subsonic or supersonic.  For $\epsilon'$ close to zero, $v\ll c_s$,
and we have subsonic conditions.  However, as
$|\epsilon'|\rightarrow2\sqrt{2}\alpha$, the radial velocity increases
and the sound speed decreases and we can have supersonic flows.

As a final comment, we recall that the rotating solutions described in
the main text of the paper exist only for $\epsilon'>C\alpha^2$ where
$C$ is a constant of order unity (\S3.1).  On the other hand, the
spherical solutions described here exist for
$\epsilon'<2\sqrt{2}\alpha$.  Thus, for
$C\alpha^2<\epsilon'<2\sqrt{2}\alpha$ both solution branches are
allowed.  What is the relationship between the two solutions?  Which
solution will an actual flow prefer?  Perhaps the answer is determined
by the initial conditions of the flow, especially the angular momentum
of the accreting gas.

\vfill\eject

\noindent{\bf References}
\bigskip

\ref{Abramowicz, M., Czerny, B., Lasota, J. P., \& Szuszkiewicz, E. 1988, ApJ,
332, 646}

\ref{Balbus, S. A. \& Hawley, J. F. 1991, ApJ, 376, 214}

\ref{Begelman, M. C. 1978, MNRAS, 184, 53}

\ref{Begelman, M. C. \& Meier, D. L. 1982, ApJ, 253, 873}

\ref{Bondi, H. 1952, MNRAS, 112, 195}

\ref{Dubrulle, B. 1992, A\&A, 266, 592}

\ref{Eggum, G. E., Coroniti, F. V., \& Katz, J. I. 1988, ApJ, 330, 142}

\ref{Frank, J., King, A., \& Raine, D. 1992, Accretion Power in Astrophysics
(Cambridge, UK: Cambridge Unniversity Press)}

\ref{Hawley, J. F., Gammie, C. F. \& Balbus, S. A. 1994, preprint}

\ref{Henriksen, R. N. \& Valls-Gaboud, D. 1994, MNRAS, 266, 681}

\ref{Landau, L. D. \& Lifshitz, E. M. 1959, Fluid Mechanics
(London: Pergamon Press)}

\ref{Liang, E. P. 1988, ApJ, 334, 339}

\ref{Lynden-Bell, D. \& Pringle, J. E. 1974, MNRAS, 168, 603}

\ref{Mihalas, D. \& Mihalas, B. W. 1984, Foundations of Radiation Hydrodynamics
(New York: Oxford University Press)}

\ref{Narayan, R., Loeb, A., \& Kumar, P. 1994, ApJ, 431, 359}

\ref{Narayan, R. \& Popham, R. 1993, Nature, 362, 820}

\ref{Narayan, R. \& Yi, I, 1994, ApJ, 428, L13}




\ref{Popham, R., Narayan, R., Hartmann, L., \& Kenyon, S. 1993, ApJ,
415, L127}

\ref{Press, W. H., Teukolsky, S. A., Vetterling, W. T., \& Flannnery, B. P.
1992, Numerical Recipes (Cambridge: Cambridge University Press)

\ref{Rees, M. J., Begelman, M. C., Blandford, R. D., \& Phinney, E. S. 1982,
Nature, 295, 17}

\ref{Ryu, D., \& Goodman, J. 1992, ApJ, 388, 438}

\ref{Shakura, N. I. \& Sunyaev, R. A. 1973, A\&A, 24, 337}

\ref{Spruit, H. C., Matsuda, T., Inoue, M., \& Sawada, K. 1987, MNRAS, 229,
517}

\vfill\eject
\noindent{\bf Figure Captions}
\bigskip\noindent
Figure 1.  Self-similar solutions corresponding to $\alpha=0.1$,
$\epsilon^{\prime}=0.1,1,10$.  Top left: angular velocity $\Omega$ as
a function of polar angle $\theta$.  Top right: radial velocity $v$.
Bottom left: density $\rho$.  Bottom right: square of the sound speed,
$c_s^2$.

\bigskip\noindent
Figure 2. Isodensity contours in the meridional plane for four
solutions.  The top two panels and the bottom left panel correspond to
the solutions shown in Figure 1.  The bottom right panel shows a
solution in which $\Omega$ reverses sign twice (see text).

\bigskip\noindent
Figure 3. Comparison of the exact solutions of this paper with the
height- integrated solutions of NY.  The solid lines correspond to the
height-integrated solutions; the dotted lines correspond to spherical
averages (cf eq. 3.5); the dashed lines correspond to cylindrical
$z$-averages (eqs. 3.6--3.8); and the long-dashed lines correspond to
midplane values.  The height-integrated values agree remarkably well
with the spherical averages for all values of $\epsilon^{\prime}$.

\bigskip\noindent
Figure 4. Dimensionless Bernoulli parameter $b$ as a function of the
spherical polar angle $\theta$ for self-similar solutions with
$\alpha=0.1$, $\epsilon=0.333$ (i.e. $\gamma=1.5$), and from bottom to
top, $\epsilon^{\prime}=300, ~100, ~10, ~1, ~0.75, ~0.33$
(i.e. $f=0.0011$, 0.0033, 0.033, 0.33, 0.4465, 1).  Note that the
Bernoulli parameter is always positive close to the rotation axis, and
is positive at all $\theta$ for $\epsilon^{\prime} <0.75$
(i.e. $f>0.446$).

\bigskip\noindent
Figure 5. The ratio between the convection term and advection term in
the energy equation, shown as a function of $\theta$, for a solution
with $\alpha=0.01$, $\alpha_c=0.005$, $\epsilon=0.01$, $f=1$.  For all
$\theta>\theta_{crit}=33^o$, the convection term is smaller than the
advection term and self-similar advection-dominated accretion is
possible.  For $\theta<\theta_{crit}$, convection dominates over
advection and we speculate that convection will initiate a bipolar
outflow in this region of the flow.  Note from Fig. 4 that this region
has the most positive Bernoulli parameter and is therefore most
susceptible to being ejected.

\bye